\documentclass[aps,prd,twocolumn,preprintnumbers,showpacs,nofootinbib]{revtex4}
\usepackage{graphics}  
\usepackage{epsfig}
 \usepackage{verbatim}

\IfFileExists{srcltx.sty}{\usepackage[active]{srcltx}}

\newcommand{\be}{\begin{equation}}
\newcommand{\ee}{\end{equation}}
\newcommand{\bea}{\begin{eqnarray}}
\newcommand{\eea}{\end{eqnarray}}

\begin{document}

\title{Gravitino dark matter from $Q$-ball decays}

\author{Ian M. Shoemaker and Alexander Kusenko}
\affiliation{Department of Physics and Astronomy, University of California, Los Angeles, CA 90095-1547, USA}
\preprint{UCLA/09/TEP/56}

\begin{abstract}
Affleck--Dine baryogenesis, accompanied by the formation and subsequent decay of Q-balls, can generate both the baryon asymmetry of the universe and dark matter in the form of gravitinos.   The gravitinos from Q-ball decay dominate over the thermally produced population if the reheat temperature $T_{R} \lesssim 10^{7} ~\rm{GeV}$.  We show that a gravitino with mass $\sim 1 ~\rm{GeV}$  is consistent with all observational bounds and can explain the baryon-to-dark-matter ratio in the gauge-mediated models of supersymmetry breaking for a wide range of cosmological and Q-ball parameters.   Moreover, decaying Q-balls can be the dominant production mechanism for $m_{3/2} < 1~\rm{GeV}$ gravitinos if the Q-balls are formed from a $(B-L)=0$ condensate, which produces no net baryon asymmetry.  Gravitinos with masses in the range $50~\rm{eV} \lesssim m_{3/2} \lesssim 100~\rm{keV}$  produced in this way can act as warm dark matter and can have observable imprint on the small-scale structure. 

\end{abstract}

\maketitle
\section{Introduction}
Supersymmetric extensions of the standard model predict the existence of baryonic and leptonic Q-balls~\cite{Kusenko:1997zq}, which may be stable, or  may decay into fermions.  In the early universe, large Q-balls are abundantly produced from the fragmentation of a flat direction condensate.  In case of gauge-mediated supersymmetry breaking, these Q-balls can be stable~\cite{Dvali:1997qv} and exist today as dark matter~\cite{Kusenko:1997si}. The cosmology and astrophysical implications of Q-balls have been studied by a number of authors~\cite{Frieman:1988ut,Frieman:1989bx,Griest:1989bq,Griest:1989cb,Enqvist:1997si,
Enqvist:1998ds,Enqvist:1998xd,Kusenko:1997ad,Kusenko:1997hj,Kusenko:1997it,Kusenko:1997vp,Laine:1998rg,Enqvist:1998en,
Enqvist:1998pf,Axenides:1999hs,Banerjee:2000mb,Battye:2000qj,Allahverdi:2002vy,Enqvist:2003gh,Dine:2003ax,Kusenko:2004yw,
Kusenko:2005du,Berkooz:2005rn, Berkooz:2005sf,Kusenko:2008zm,Kusenko:2009cv,
Johnson:2008se,Kasuya:2008xp,Sakai:2007ft,Campanelli:2007um,Kasuya:2000wx,
Kawasaki:2005xc,Kasuya:2007cy,Shoemaker:2008gs,Campanelli:2009su,Kusenko:2009iz,Shoemaker:2009ru}.  Here we study the cosmological implications of small-charge, unstable Q-balls, which may not exist at present, but whose formation and decay in the early universe could produce the population of dark matter particles consistent with observations. 

{\em A priori}, one could expect the baryon-to-dark matter (BDM) ratio, $\Omega_{b}/\Omega_{DM}$ to be very different from one if dark matter and baryonic matter arise from very different processes. The observational fact that $\Omega_{b}/\Omega_{DM} = 0.17$, is within one order of magnitude from unity is intriguing.  This may be an indication that baryons and dark matter share a common physical origin~\cite{Kaplan:1991ah}. Most mechanisms of dark matter production are completely unrelated to baryogenesis and therefore implicitly assume that this ratio is an accident.  In the case of Affleck-Dine cosmology~\cite{Affleck:1984fy,Dine:2003ax}, dark matter can originate from the same process as baryogenesis, which suggests a possible explanation to the fact that $\Omega_{b}$ and $\Omega_{DM}$ are not drastically different.   Models along these lines have been constructed for both gravity-mediated and gauge-mediated supersymmetry breaking models. In gauge-mediated models the partial evaporation of baryons off the Q-ball are the source of baryons, while the remaining Q-balls act as dark matter~\cite{Laine:1998rg,Banerjee:2000mb,Kasuya:2001hg}. In gravity-mediated models the late decay of Q-balls produces baryons and non-thermal LSP dark matter~\cite{Enqvist:1997si,Enqvist:1998en}. 
	
We will show that decays of unstable Q-balls in the gauge-mediated scenario can serve as the dominant and non-thermal source of dark matter in the form of gravitinos, hence relating dark matter and baryon densities in a new way.  In our model a $\sim 1~\rm{GeV}$ gravitino is favored. In conventional scenarios, the gravitinos are produced in a combination of thermal processes and NLSP decays, $m_{3/2} \sim 1~\rm{GeV}$.  Such a population of gravitinos can be dark matter, but the scenario requires a finely tuned reheating temperature, $10^{6}~\rm{GeV} \lesssim T_{R} \lesssim 10^{7}~\rm{GeV}$ \cite{Steffen:2006hw}.  In models of gauge-mediated SUSY breaking in which the flat direction dynamics are properly taken into account, the reheat temperature is often significantly below this desired value~\cite{Allahverdi:2005mz,Allahverdi:2007zz}.  Our production mechanism extends the viability of gravitino dark matter to low reheating temperature. 

While a somewhat similar model of gravitino production from Q-ball decays has been studied before in the context of gravity-mediated SUSY breaking~\cite{Seto:2005pj}, the $\rm{GeV}$ scale of gravitino mass is theoretically disfavored in gravity-mediated scenarios.  Moreover, the model considered in Ref.~\cite{Seto:2005pj} required a high reheating temperature and neglected some important thermal processes in calculating the Q-ball decay temperature. 

Enqvist and McDonald~\cite{Enqvist:1997si,Enqvist:1998en} have discovered a very simple way of creating the correct amounts of dark matter and baryon number simultaneously.  The crucial observation is that the squarks $\tilde{q}$ in unstable Q-balls decay into quarks $q$ and the LSP via $\tilde{q} \rightarrow \rm{LSP} + q$.  For every baryon number lost by the Q-ball at least three LSPs are created, $N_{LSP} \ge 3$.  Since Q-ball formation takes place non-adiabatically the fraction of baryon number trapped in the Q-balls, $f_{B}$, is generally maximal $f_{B} \sim 1$~\cite{Kasuya:1999wu}.  Then one finds  
\be \label{ratio} \frac{ \Omega_{b}}{\Omega_{LSP}} = \frac{ m_{n} Y_{b}}{m_{LSP} Y_{LSP}} = \frac{m_{n}}{m_{LSP}} \frac{1}{f_{B} N_{LSP}}. \ee
Since Q-balls are a large conglomeration of squarks, it should not be surprising that in an $R$-parity preserving theory their decay necessarily creates comparable amounts of LSPs and baryons. In the original model, Enqvist and McDonald \cite{Enqvist:1997si,Enqvist:1998en} considered gravity-mediation, in which the LSP dark matter was the neutralino. However a $\rm{GeV}$ neutralino is theoretically disfavored and now experimentally ruled out by LEP. To maintain the prediction of eq.~(\ref{ratio}),  Roszkowski and Seto~\cite{Roszkowski:2006kw} extended this idea to the case of axino LSP in gravity mediation which naturally accounts for the $\rm{GeV}$ scale.  However the model is most natural in the gauge-mediated case, which already has a light LSP, namely, the gravitino.  This is intriguing because gravitinos with $\mathcal{O}(1~\rm{GeV})$ masses are not only natural in gauge-mediated supersymmetry models, but this scale is particularly attractive in its ability to solve several cosmological and phenomenological problems \cite{Ibe:2007km,Feng:2008zza}.  Moreover, although large gauge-mediated Q-balls can be stable, they are unstable for $Q \lesssim 10^{12}$ at zero temperature and evaporate by the present-day time from thermal effects for $Q \lesssim 10^{18}$ \cite{Kasuya:2001hg}.  It is these Q-balls we focus on as the source of the correct BDM ratio.  Moreover, decaying Q-balls can be the dominant production mechanism for $m_{3/2} < 1~\rm{GeV}$ gravitinos if the Q-balls are formed from a $(B-L)=0$ condensate which produces no baryons. In this case of course the novel feature is a new mechanism of gravitino production, and the BDM ratio cannot be accounted for.

\section{Reheating temperature} 

To avoid gravitino overclosure, the reheating temperature must not exceed the following upper bound~\cite{Moroi:1993mb,deGouvea:1997tn}: 
\be T_{R} \lesssim \left(10~ \rm{TeV}\right) h^{2}~ \left(\frac{m_{3/2}}{100~\rm{keV}}\right) \left(\frac{\rm{TeV}}{M_{G_{3}}}\right)^{2}, \ee 
where $M_{G_{3}}$ is the gluino mass. For a $\rm{GeV}$ gravitino the upper bound on the reheating temperature is $10^{8}~\rm{GeV}$. A more stringent constraint however comes from requiring that the late decays of the NLSP not disrupt successful big bang nucleosynthesis: $T_{R} \lesssim 10^{7}~\rm{GeV}$ \cite{Gherghetta:1998tq, Giudice:1998bp}.  The reheat temperature in the case of stop or sneutrino NLSP is less constrained than in the case of stau NSLP:
\be 
  T_{R} \lesssim \left\{
     \begin{array}{l}
 10^{7}~\rm{GeV} , \ \ \  \rm{for \ NLSP} = \tilde{\tau}\\
         10^{8}~\rm{GeV},  \ \ \  \rm{for \ NLSP} = \tilde{t}, \tilde{\nu}.\\
     \end{array}
   \right.
   \label{zen}
\ee
\section{Gravitino abundance from Q-ball decays}
Gravitinos freeze out at a very high temperature \cite{deGouvea:1997tn,Steffen:2006hw}
\be T_{f} \approx 200~ \rm{TeV} \left(\frac{m_{3/2}}{100~\rm{keV}}\right)^{2} \left(\frac{1~\rm{TeV}}{M_{G_{3}}}\right)^{2}. \ee
For the $1~\rm{GeV}$ gravitinos with $\rm{TeV}$-scale gluinos the freeze-out temperature is $T_{f} \sim 2 \times 10^{13}~\rm{GeV}$. As long as Q-balls decay well below this temperature, the produced gravitinos do not thermalize.  Notice that we must have $T_{R} \ll T_{f}$ to avoid gravitino overclosure. However, gravitinos are produced from scatterings in the hot plasma of MSSM particles.  

We now compute the Q-ball decay temperature and show that the produced gravitinos do not thermalize.  At finite temperature the effective potential along a flat direction is \cite{Kusenko:1997si,Laine:1998rg,Kasuya:2001hg}
\be V(\phi, T) = m^{4}(T)~ \log \left( 1+ \frac{|\phi|^{2}}{m^{2}(T)}\right), \ee
where $m(T) = \rm{max} \left(M_{s},T \right)$, $M_{S}$ is the supersymmetry breaking scale, and $T$ is the temperature. The mass of a squark inside the Q-ball is $\omega \sim m(T)/Q^{1/4}$.  Notice then that while all Q-balls with $Q\lesssim 10^{12}$ are unstable at low-temperature ($T < M_{S} \sim 1~\rm{TeV}$), larger Q-balls may decay as well via $\tilde{q} \rightarrow \tilde{G} + q$ at high temperature. 

At finite temperature, both thermal evaporation of Q-balls and accretion of global charge onto Q-balls are possible~\cite{Frieman:1988ut,Griest:1989bq,Kusenko:1997hj,Laine:1998rg,Banerjee:2000mb,Kasuya:2001hg}.  The fate of Q-balls is, to some extent, determined by the difference between the chemical potential of the surrounding plasma $\mu_{\rm plasma}$ and inside the Q-ball $\mu_{Q}\sim \omega $.  For sufficiently small charge Q-balls at high temperature, the Q-ball chemical potential dominates $\mu_{Q} \gg \mu_{plasma}$, thus leading to the evaporation of such Q-balls \cite{Laine:1998rg}, so that all Q-balls with charges $Q\lesssim 10^{18}$ would have evaporated by the present era \cite{Kasuya:2001hg}. 
 
At high temperatures the effects of diffusion are important, while at lower temperatures thermal evaporation dominates. The temperature at which this transition occurs is  $T_{*}\approx M_{s} \approx 1~ \rm{TeV}$ for small charge Q-balls.
Kasuya and Kawasaki \cite{Kasuya:2001hg} have shown that at early times ($T > T_{R}$), the charge transfer rate is 
\be \frac{dQ}{dT} \sim 10 \frac{M T_{R}^{2}}{T^{4}}. \ee
Thus the charge evaporated from a Q-ball during this epoch is 
\be \Delta Q = 3M T_{R}^{2} \left(\frac{1}{T_{R}^{3}}-\frac{1}{T_{i}^{3}} \right) \approx \frac{3M}{T_{R}}, \ee
where we have evolved the temperature from $T_{i}$ to $T_{R}$, assuming $T_{i} \gg T_{R}$.  Thus for $T_{R} \lesssim 10^{7}~\rm{GeV}$ Q-balls with charges $Q\gtrsim 10^{11}$ survive until after reheating. 

Q-balls decaying before reheating decay at the temperature ($T \gtrsim T_{R}$)
\be T_{D} \approx 10^{7}~ \rm{GeV} \left(\frac{10^{11}}{Q}\right)^{1/3} \left(\frac{T_{R}}{10^{7}~\rm{GeV}}\right)^{2/3}. \ee
Since the Q-balls evaporated in this era are small $Q\lesssim 10^{11}$ and the $T_{R} \lesssim 10^{7}~\rm{GeV}$, the decay temperature $T_{D} \ll T_{f}$, which ensures that the gravitinos produced do not thermalize. 

The Q-balls in the range $10^{11} \lesssim Q \lesssim 10^{18}$ can survive until well after reheating. If they decay when diffusion is still dominant, then their decay temperature is ($T_{*} \lesssim T_{D} \lesssim T_{R}$)
\be T_{D} \sim \frac{10 M}{Q} \sim 10^{3}~\rm{GeV} \left(\frac{10^{16}}{Q}\right). \ee
The largest Q-ball that decays in this regime is $Q\sim 10^{16}$.  One can show that the Q-balls in the range $10^{16} \lesssim Q \lesssim 10^{18}$ all decay very close to $T_{D} \approx 10^{2}~\rm{GeV}$.  

Thus the Q-balls in the range $10^{12} \lesssim Q \lesssim 10^{18}$ decay after reheating, producing baryons and non-thermal gravitinos. The gravitino abundance comes potentially form three sources: thermal production, non-thermal production from the NLSP decays, and from Q-ball decay
\be Y_{G} = Y_{G}^{TP} + Y_{G}^{NLSP} + Y_{G}^{Q}. \ee
However in the case of charged slepton NLSP, their decay contributions to the gravitinos is subdominant compared to thermal processes for $m_{3/2} \lesssim 10~\rm{GeV}$  \cite{Steffen:2006hw}. We therefore focus on the comparison between $Y_{G}^{Q}$ and $Y_{G}^{TP}$. The bound on the reheating temperature already ensures that the first two terms will not lead to overclosure.  The number density owing to Q-ball decays is 
\be Y_{G}^{Q} = N_{G} f_{B} \frac{n_{b}}{s}. \label{qprod} \ee
Using $n_{b}/s \approx 5\times10^{-10}$ with  $N_{G} \approx 3$ and $f_{B} \approx 1$, we have $Y_{G}^{Q} \sim 10^{-9}$. The abundance from thermal processes is \cite{deGouvea:1997tn,Steffen:2006hw}
\be Y_{G}^{TP} = 3\times 10^{-12}  \left(\frac{1~\rm{GeV}}{m_{3/2}}\right)^{2} \left(\frac{T_{R}}{10~\rm{TeV}}\right). \ee
Thus for a $\rm{GeV}$ gravitino the Q-ball production mechanism is dominant ($Y_{G}^{Q} \gtrsim Y _{G}^{TP}$) as long as $T_{R} \lesssim 10^{7} \rm{GeV}$. This automatically guarantees that the gravitino and the NLSP do not affect the successful predictions of conventional cosmology.  Gravitinos produced from either thermal processes or Q-ball decay extend the viability of gravitino dark matter to reheating temperatures $T_{R} \lesssim 10^{7} \rm{GeV}$, though only when Q-ball decay is dominant can the BDM ratio be successfully accounted for. Additionally the dependence on the specific flat direction is weak, although we need $(B-L) \neq 0$ to prevent sphalerons from washing-out any of the produced $B$ and $L$ asymmetry. 

It has been pointed out \cite{Pospelov:2006sc} that long-lived, negatively charged electroweak-scale particles $X^{-}$ can overproduce $\rm{Li}^{6}$ via the reaction $\left(^{4}\rm{He} X^{-1}\right) + \rm{D} \rightarrow \rm{Li}^{6} + X^{-}$. Such overproduction of $\rm{Li}^{6}$ constrains the lifetime of the $X^{-}$ to be $\tau \lesssim 5 \times 10^{3}~ \rm{s}$. This constraint is directly relevant to gauge-mediated models in which charged sleptons are often the NLSP. However we note that the NLSP lifetime is given by 
\be \tau_{\tau} = \frac{48\pi m_{3/2}^{2}M_{P}^{2}}{m_{\tau}^{5}} \left( 1 - \frac{m_{3/2}^{2}}{m_{\tau}^{2}}\right)^{-4},\ee
where $m_{\tau}$ is the stau mass. For $\rm{GeV}$-scale gravitinos this implies the very weak bound $m_{\tau} \gtrsim 60~\rm{GeV}$.  Thus no significant constraints from BBN are imposed on $\rm{GeV}$ gravitinos.  

\section{Beyond $\rm{GeV}$-scale gravitinos}
Even if we abandon the desire to account for $(\Omega_{b}/\Omega_{DM})$, the gravitinos produced from Q-ball decay can dominate other production mechanisms for $m_{3/2} \lesssim \rm{GeV}$ and $T_{R} \lesssim 10^{7} ~\rm{GeV}$.  Since in this case we are not attempting to explain $\eta_{B} \sim 10^{-10}$, the Q-balls must be formed from a $(B-L)=0$ flat direction, since sphalerons will destroy any baryon asymmetry. The gravitino yield is still described by Eq. \ref{qprod}, but now $n_{b}$ is replaced by $n_{(B-L)}$ which can be much larger  $\eta_{(B-L)} \gg 10^{-10}$~\cite{Kusenko:2008zm,Kusenko:2009cv}.  This feature allows the gravitino yield from Q-ball decays to  exceed the yield from other sources. Now however there is a lower-bound on the decay temperature of the Q-balls, $T_{D} > T_{ew}$, so that the produced baryon/lepton asymmetries are erased. As we have seen, however, this constraint is not difficult to satisfy. 

\section{Free-Streaming Length}

Although the $\rm{GeV}$-mass gravitinos from Q-ball decays constitute a cold dark matter candidate, smaller mass gravitinos can have a cosmologically relevant free-streaming length.   One can estimate the free-streaming length by calculating the present-day free-streaming velocity $v_{0}$ \cite{Steffen:2006hw,Jedamzik:2005sx}

\bea \lambda_{FS}^{G} & &\approx 2v_{0} ~t_{eq}~ (1+ z_{eq})^{2} \nonumber \\ & & \times \log \left( \sqrt{ 1+ \frac{1}{v_{0}^{2}(1+z_{eq})^{2}}}+ \frac{1}{v_{0}(1+z_{eq})}\right). \eea

The free-streaming length limits from the Lyman-$\alpha$ forrest constrain $\lambda_{FS} \lesssim 0.59~\rm{Mpc}$ \cite{Narayanan:2000tp,Steffen:2006hw}, which translates to the limit $v_{0} \lesssim 0.05 ~\rm{km/s}$ on the free-streaming velocity. At the time of decay the momentum of the gravitino is roughly 
\be p_{G} = \frac{ \omega^{2} - m_{3/2}^{2} - m_{p}^{2}}{2 \omega}, \ee
which yields the free-streaming velocity 
\be v_{0}^{G} = \frac{T_{0} }{2 Q^{1/4} m_{3/2}} \left( \frac{g_{0}}{g_{d}}\right)^{1/3} \left[1- \frac{2 m_{p}^{2} Q^{1/2}}{T_{D}^{2}}\right], \ee
where $T_{0}$ is the present-day temperature, $T_{D}$ is the Q-ball decay temperature and $m_{p}$ is the proton mass. 

We note that since small $Q$ translates into high decay temperature (early times), the small Q-balls have a chance of producing warm dark matter. The earlier produced gravitinos that are warmer.  For the $\rm{GeV}$-scale gravitino, the present-day free-steaming velocity is $\lesssim 10^{-9}~ \rm{km/s}$, which corresponds to conventional cold dark matter.   

The gravitinos with masses less than $\rm{GeV}$ may  have a cosmologically relevant free-streaming length.  In Fig.~1 we plot the limits on the Q-ball produced gravitinos for $\lambda_{FS} = 1~\rm{Mpc}$, $100~\rm{kpc}$, $10~\rm{kpc}$, and $1~\rm{kpc}$. There are different limits on $\lambda_{FS}$ from different observations. There are limits based on the Lyman-$\alpha$ forrest at $z=3$ ($\lambda_{FS} \lesssim 0.59~\rm{Mpc}$) \cite{Narayanan:2000tp}, while reionization has very different limits ($\lambda_{FS} \lesssim 0.94~\rm{Mpc}$) \cite{Barkana:2001gr}. 

It has been argued in fact, that thermal gravitino dark matter is now ruled out for $m_{3/2} < 16~\rm{eV}$ based on Lyman-$\alpha$ limits at $z \approx 2-3$ \cite{Viel:2005qj}.  However this constraint is only directly applicable to the thermally produced gravitino dark matter. If instead the reheating temperature is small $T_{R} \lesssim 10^{7}~\rm{GeV}$ as we have argued here, then the results of \cite{Viel:2005qj} are evaded.  A number of similar proposals for evading Lyman-$\alpha$ bounds exist. For example, entropy production from messenger decay \cite{Baltz:2001rq} and the late decay of the relic neutralino population both produce gravitinos that are cool enough to evade Lyman-$\alpha$ constraints \cite{Feng:2008zza}.  Both of these solutions however prefer a particular gravitino mass: $1~\rm{keV}$ in the case of messenger entropy production \cite{Baltz:2001rq}; and $m_{3/2} \sim \left(1-10\right)\rm{GeV}$ from late neutralino decay  \cite{Feng:2008zza}. As can be seen from Fig.~$1$ however, Q-ball produced gravitinos are viable over a much wider range of masses:  $50~\rm{eV} \lesssim m_{3/2} \lesssim 100~\rm{keV}$. 

Moreover, observations of the motions of stars in dwarf spheroidal galaxies show evidence of a non-zero free-streaming length \cite{Wyse:2007zw}. Observations therefore seem to prefer a free-streaming length close to the Lyman-$\alpha$ limits. Thus for a relatively wide range of parameters Q-ball decay can produce phenomenologically viable warm dark matter.

\begin{figure}
\begin{center}
\includegraphics[scale=0.5]{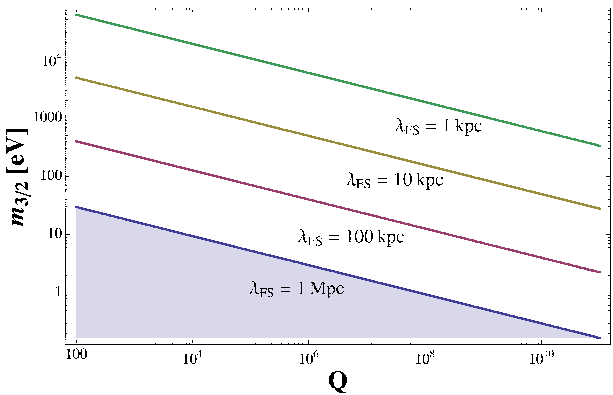}
\caption{Here we show the parametric dependence of the free-streaming length $\lambda_{FS}$ on Q-ball charge $Q$ and gravitino mass $m_{3/2}$.  The blue region is excluded because it gives rise to hot dark matter with $\lambda_{FS} \gtrsim 1~\rm{Mpc}$. }
\end{center}
\label{figure1}
\end{figure}

\section{Conclusions} 
As long as $T_{R} \lesssim 10^{7}~\rm{GeV}$, Q-balls with baryon numbers $10^{12}-10^{18}$ decay after reheating and preserve the successful Enqvist-McDonald relation relating baryonic and dark matter energy densities 
\be \left(\frac{\Omega_{3/2}h^{2}}{0.11}\right) \approx \left(\frac{m_{3/2}}{1~\rm{GeV}}\right) \left(\frac{N_{G}f_{B}}{3}\right) \left(\frac{\Omega_{b} h^{2}}{0.02}\right). \ee
Here we took $N_{G} =3$ and $f_{B} \approx 1$, consistent with numerical simulations~\cite{Kasuya:1999wu}. 
Thus the gravitino mass consistent with the correct BDM ratio is

\be \frac{1.6~\rm{GeV}}{f_{B}} \lesssim ~m_{3/2}~ \lesssim \frac{2.0~\rm{GeV}}{f_{B}}. \ee 

It is instructive to compare the production of gravitinos from Q-ball decays, the thermal freeze-out processes, and the non-thermal late decays of NLSPs.  In particular, the three possibilities have different implications for the reheat temperature.   If the thermally produced population dominates the relic abundance, then the reheating temperature must be finely-tuned to give $\Omega_{DM} h^{2} \approx 0.11$.  If the contribution from NLSP decay dominates, then the reheating temperature must be relatively small $T_{R} \ll 10^{9}~\rm{GeV}$ and generally requires a larger gravitino mass $m_{3/2} \gtrsim 10~\rm{GeV}$.  Finally, the production from Q-ball decays requires a small reheating temperature $T_{R} \lesssim 10^{7}~\rm{GeV}$ and small gravitino mass $m_{3/2} \lesssim 1~\rm{GeV}$. In the special case with the gravitino mass $m_{3/2} \sim 1~\rm{GeV}$ one can account for the correct  baryon-to-dark matter ratio.  

Finally, if the Q-balls form from a $(B-L)=0$ Affleck-Dine condenstate({\em cf.} the models in Refs.~\cite{Kusenko:2008zm,Kusenko:2009cv}), the decays of Q-balls produce no net baryon asymmetry that can survive the sphaleron transitions, but these decays can still act as a source of non-thermal gravitinos.  In this case, if the gravitino mass is in the range $50~\rm{eV} \lesssim m_{3/2} \lesssim 100~\rm{keV}$ Q-ball decay produces dark matter with a cosmologically interesting free-streaming length that can be tested in astronomical observations~\cite{Wyse:2007zw,Coe:2009wt}.  

This work was supported in part  by DOE grant DE-FG03-91ER40662 and by the
NASA grant  NNX08AL48G.  

\bibliography{qball}
\bibliographystyle{h-physrev3}

\end{document}